\documentclass[aps,prab,preprint]{revtex4-1}  %% For PRAB
\usepackage{graphicx}
\usepackage{amsmath}
\usepackage{amssymb}
\usepackage{amsbsy}
\usepackage{bm}   % for bold math fonts, especially greek
\newcommand{\bit}{\begin{itemize}}
\newcommand{\eit}{\end{itemize}}

\def\benu{\begin{enumerate}}
\def\eenu{\end{enumerate}}
\def\noi{\noindent}
\def\btab{\begin{tabbing}}
\def\etab{\end{tabbing}}

\def\bit{\begin{itemize}}
\def\eit{\end{itemize}}
\def\beq{\begin{equation}}
\def\eeq{\end{equation}}
\def\bec{\begin{center}}
\def\eec{\end{center}}
\def\btable{\begin{tabular}}
\def\etable{\end{tabular}}
\def\beqr{\begin{eqnarray}}
\def\eeqr{\end{eqnarray}}
\def\rarw{\rightarrow}

\def\om{\omega}
\def\gm{\gamma}

\def\eps{\epsilon}

\def\al{\alpha}
\def\bt{\beta}

\def\Dl{\Delta}
\def\sg{\sigma}

\def\rarw{\rightarrow}
\def\del{\partial}

\def\half{\frac{1}{2}}

\def\btab{\begin{tabbing}}
\def\etab{\end{tabbing}}
\def\beqrs{\begin{eqnarray*}}
\def\eeqrs{\end{eqnarray*}}
\def\noi{\noindent}
\def\lan{\langle}
\def\ran{\rangle}
\def\bfig
{\begin{figure}}
\def\efig{\end{figure}}

 \fontfamily{ptm}\selectfont
 \usepackage{mathptmx}           % this gives slanted Times symbols for math

\begin{document}

\title{Diffusion measurement from observed transverse beam echoes}

1%%% With revtex
\author{Tanaji Sen}
\affiliation{Accelerator Physics Center, FNAL, Batavia, IL 60510}
\email{tsen@fnal.gov}
\author{Wolfram Fischer}
\affiliation{BNL, Upton, NY 11973}
%\date{}

\begin{abstract}
We study the measurement of transverse diffusion through beam echoes. We revisit earlier observations of echoes in RHIC and apply an updated theoretical model to these measurements.
We consider three possible models for the diffusion coefficient and show that only one is 
consistent with measured echo amplitudes and pulse widths. This model allows us to 
parameterize the diffusion coefficients as functions of bunch charge. 
We demonstrate that echoes can be used to measure diffusion much quicker
than present methods and could be useful to a variety of hadron
synchrotrons. 
\end{abstract}

\maketitle

%\tableofcontents

\section{Introduction}
Beam diffusion can lead to emittance growth, halo formation and particle loss.
A standard method currently used to measure transverse diffusion requires scraping the beam with 
collimator jaws moved close to the beam, then retracting the jaws and waiting for the beam 
to diffuse to the outer
position of the jaws \cite{Mess_Seidel, Fischer_97, Fliller, Valentino, Stancari}. This procedure is time consuming and the 
method is only applicable to storage rings where the beam circulates for times long enough 
to enable the measurement. Beam echoes were introduced into accelerator physics more than 
two decades ago \cite{Stupakov, Stup_Kauf} and then shown to be useful as a novel method to 
measure transverse diffusion \cite{Stup_Chao}.
A single echo observation can be done typically within a thousand turns with 
nonlinear tune spreads in the range 0.001 - 0.01. Hence diffusion measurements with echoes
would be considerably faster than the standard method and could also enable diffusion 
to be measured in synchrotrons where beams circulate for relatively short times.

Shortly after the introduction of the beam echo concept, longitudinal unbunched beam echoes 
were observed at the Fermilab Antiproton
Accumulator \cite{Fermi_AA} and then at the CERN SPS \cite{CERN_SPS} 
The original motivation however had been to
measure transverse diffusion from transverse echoes. In the year 2000, transverse 
bunched beam echoes were observed in the SPS with two consecutive dipole kicks
\cite{Arduini} but no diffusion coefficients were extracted. Later in 2004-2005 an 
extensive set of dedicated experiments was carried out at RHIC with dipole and
quadrupole kicks
\cite{Fischer_PAC05} and these will be the focus in this paper. The existing model 
as applied to the data did not yield consistent values for the diffusion coefficients \cite{Sorge}. 

The next generation of intensity frontier hadron synchrotrons will require tight control of
particle amplitude growth. At Fermilab the Integrable Optics
Test Accelerator (IOTA) \cite{IOTA} ring is under construction where the novel concept of 
nonlinearly integrable lattices will be tested and could serve as a model for future 
synchrotrons. This ring offers the
opportunity of testing a fast diffusion measurement technique which could help determine the degree of
integrability (or stable motion) among different lattice models. With this motivation, 
we revisit the earlier RHIC measurements with an updated theoretical model to enable 
extraction of self-consistent diffusion coefficients. In Section II we describe the 
updated model, in Section III we apply this model to the RHIC data, in Section IV we 
consider beam related time scales and we summarize in Section V with lessons to be applied to
future echo measurements.

\section{Echo pulse with diffusion}

The basic beam echo generating mechanism is well known. If at some initial
time the beam is kicked away from the central orbit, the beam centroid
will decohere due to a nonlinear spread of frequencies. If subsequently
a quadrupole kick is applied after the centroid response has decayed away,
a diminished coherent response will reappear after a time interval equal
to the delay between the dipole and quadrupole kicks. 
Figure \ref{fig: echopulse_1_fwhm} in Section \ref{subsec: amp_fwhm} shows 
an example of this echo formation during the measurements at RHIC. 

Here we discuss the model to calculate the echo amplitude with diffusion using the same method and 
notation as in \cite{Chao}. 
The phase space coordinates used $x, p$ and action angle coordinates $J, \phi$ are related as
\beqr
x = \sqrt{2\bt J}\cos\phi , \; \; \; & \;\;\; p = \al x + \bt x' = - \sqrt{2\bt J} \sin\phi  \\
J = \frac{1}{2\bt}(x^2 + p^2)  ,   &   \tan\phi = -\frac{p}{x} 
\eeqr
The initial distribution is taken to be exponential in the action
\beq
\psi_0(J) = \frac{1}{2\pi J_0}\exp[ - \frac{J}{J_0}]
\eeq
where $J_0 = \eps_0$, the initial rms emittance. 

We first consider the dipole moment after a dipole kick and the general case where the dipole kicker is at a non-zero phase advance from the BPM 
location where the centroid is measured. 
Following the procedure in \cite{Chao}, the dipole moment after the dipole kick by an angle $\theta$ is
\beq
\lan x\ran^{amp}(t)  =\frac{ \theta  \sqrt{\bt_K \bt}}{(1+\Theta^2)}\exp[-\frac{\bt_K\theta^2}{2J_0} \frac{\Theta^2}{1+\Theta^2}]
\eeq
where $\bt_K, \bt$ are the beta functions at the kicker  and BPM respectively, 
$\Theta = \om' J_0 t$ with $\om' \equiv d\om/dJ$ the constant slope of the
 betatron angular frequency with action. This 
moment is independent of the phase advance from the kicker to the BPM. It differs from the 
expression in \cite{Chao} only by the replacement of $\bt$ by the geometric mean 
$\bt_G = \sqrt{\bt_K \bt}$ and $\bt$ in the exponent replaced by $\bt_K$. Following the 
dipole kick, the beam decoheres with the centroid amplitude decaying over a characteristic
time $\tau_D = 1/(\om' J_0)$, the decoherence time. 
At time $\tau \gg \tau_D$ after the dipole kick, a single turn quadrupole kick
is applied to generate the echoes, the first of which occurs around time 2$\tau$. The echo 
amplitude and pulse shape is affected by the diffusive beam motion. 
We consider the density distribution to evolve according to the conventional form of the diffusion
equation
\beq
\frac{\del}{\del t}\psi = \frac{\del}{\del J}[D(J)\frac{\del}{\del J}]\psi
\eeq
Here the diffusion coefficient $D(J)$ has the usual dimension of 
[action$^2$/time] and it 
differs from the definition of $D(J)$ used in \cite{Stup_Chao, Chao}. 
The treatment in \cite{Chao} had developed the theory of the echo response to first order in the
quadrupole kick strength. Since the
experiments reported in \cite{Fischer_PAC05} had observed a linear increase of the echo amplitude 
with quadrupole strength, this theory should suffice to discuss these 
experiments. 
We note that the theory developed earlier in \cite{Stup_Kauf} was nonlinear in this strength 
parameter. Using the method of \cite{Chao}, we find that the echo amplitude 
near time $t > 2\tau$ is
\beqr
\lan x \ran(t)
 & = & -\pi \bt_K \theta q \tau\int dJ \om' J^2 \psi_0'\exp[-\frac{1}{3}D(J)(\om')^2 t_1^3]
\sin(\om(t - 2\tau))
\eeqr
where $q$  is the dimensionless quadrupole kick strength defined as $q = \bt_Q/f$, the ratio of the beta function at the quadrupole to its focal length and we defined $ t_1^3 = (t-\tau)^3 + \tau^3 $.
We consider the action dependent transverse angular frequency to be of the form
$\om(J) = \om_{\bt} + \om' J$ where $\om_{\bt}$ is the angular betatron frequency 
and we consider the diffusion coefficient to be of the form
\beq
D(J) = \sum_{n=0} D_n (\frac{J}{J_0})^n
\eeq
where all coefficients $D_n$ have the same dimensions. 
The average dipole moment is given by
\beqr
\lan x\ran(t)  & = & \half \bt_K \theta q \mu \tau \om_{rev} \exp[ -\frac{1}{3}D_0 (\om')^2 t_1^3] 
{\rm Im}[e^{[i \Phi_0]} \int_0^{\infty} z^2 \exp[- z - \frac{1}{3}(\om')^2 t_1^3 \sum_n D_n z^n]
e^{[i\Phi_1 J_0 z]} dz  \nonumber \\
 \mbox{} 
\label{eq: dipmoment_n}
\eeqr
where $\om_{rev}$ is the angular revolution frequency, $\Phi_0 = \om_{\bt}(t - 2\tau)$ and
$\Phi_1 = \om'(t - 2\tau)$.
Using $ \om' = (\om_{rev}/\eps)\mu $ where $\mu= \nu(\eps) - \nu_{\bt}$ is 
the 
tune shift (from the bare tune $\nu_{\bt}$) at an action equal to the emittance, it is convenient to define scaled 
diffusion coefficients $d_n$ as
\beq
d_n =  \frac{2}{3}D_n (\frac{\om_{rev}}{\eps})^2
\label{eq: dn_Dn}
\eeq
These coefficients $d_n$ have the dimension of ${\rm time}^{-3}$. 
In the following we will consider specific cases of the above general form of $D(J)$.

Different physical processes contribute to the diffusion coefficients $D_n$. It is likely
that space charge effects, beam-beam interactions (not present in the RHIC measurements
discussed below) and intra-beam scattering all contribute to $D_0$ and higher order
coefficients. Early studies at the Tevatron at injection energy \cite{Chen_92} with additional 
sextupoles as the driving nonlinearity had measured a constant $D_0$ term which varied with the
proximity to a fifth order resonance. 
 Measurements at the LHC at top energy during collisions
showed that diffusion at the smallest amplitude measurable was finite \cite{Valentino},
implying a non-zero $D_0$. A numerical simulation \cite{Zimmermann} showed that modulation 
diffusion leads to a constant diffusion term. Beam-gas scattering and noise in dipoles 
lead to a $D_1$ term while noise in quadrupoles leads to a $D_2$ term. There are likely
other sources for these coefficients. Given that the beam is subject to
multiple effects, the complete action dependence of the diffusion may be complex. Here we
focus on the three simplest models with two diffusion coefficients that can be compared to
measurements. 

In the first case, we assume that the diffusion is of the form
\beq
D(J) = D_0 + D_1(\frac{J}{J_0})
\eeq
in this case, the dipole moment is given by
\beqr
\lan x \ran (t) & = & \bt_K \theta q \om' \tau J_0 \exp[-\frac{1}{2}d_0 \mu^2 t_1^3]
 \frac{[ (3 \al^2 - \xi^2)\xi\cos\Phi_0 + (\al^2 - 3 \xi^2)\al \sin\Phi_0]}{(\al^2 + \xi^2)^3} 
\label{eq: amp_0_1} \\
 t_1^3 & = & (t-\tau)^3 + \tau^3, \;\; \Phi_0 = \om_{\bt}(t - 2\tau),  \;\; 
\al  = 1 + \frac{1}{2}d_1 \mu^2 t_1^3,  \;\;  \xi = \om_{rev}\mu (t - 2\tau)  \nonumber 
\eeqr

The second case is  the quadratic dependence model where
\beq
D(J) = D_0 + D_2(\frac{J}{J_0})^2
\eeq
The general time dependent form of the echo at time $t=2\tau + \Dl t$ where $\Dl t $ can have either
sign is
\beqr
\lan x(t) \ran^{amp} & = &  \half \bt_K \theta q \om_{rev}\mu \tau \exp[ - \frac{1}{2}d_0 \mu^2 t_1^3]
{\rm Im}[e^{i \Phi_0} H_{02}] \label{eq: amp_0_2} \\
H_{02}(\Dl t) & \equiv & \int_0^{\infty} z^2 \exp[- a_0 z - b_2 z^2] dz \nonumber \\
 & = &  \frac{1}{8}(\frac{1}{b_2} )^{5/2}\left\{
 \sqrt{\pi}\left[a_0^2 + 2 b_2\right]\exp(\frac{a_0^2}{4b_2})
{\rm Erfc}(\frac{a_0}{\sqrt{2 b_2}}) - a_0 \sqrt{2b_2}\right\}
\nonumber \\
a_0 & = & 1 - i \xi = 1 - i\om_{rev}\mu\Dl t, \;\;\; b_2  = \half d_2 \mu^2 t_1^3 =  
\half d_2 \mu^2 [(\tau + \Dl t)^3 + \tau^3] \nonumber \\
\eeqr
Here Erfc is the complementary error function.

The last case we consider is the linear and quadratic dependence
\beq
D(J) =  D_1(\frac{J}{J_0}) +  D_2(\frac{J}{J_0})^2
\eeq
In this case, the time dependent form of the echo at time $t=2\tau + \Dl t$ is
\beqr
\lan x(t) \ran^{amp} & = & \half \bt_K \theta q \om_{rev} \mu \tau {\rm Im}[e^{i \Phi_0} H_{12}(\Dl t)]   \label{eq: amp_d1_d2}  \\
H_{12}(\Dl t) & \equiv & \int_0^{\infty} z^2 \exp[- a_1 z -  b_2 z^2] dz \nonumber \\
 & = &  \frac{1}{8}(\frac{1}{b_2} )^{5/2}\left\{
 \sqrt{\pi}\left[a_1^2 + 2 b_2\right]\exp(\frac{a_1^2}{4 b_2})
{\rm Erfc}(\frac{a_1}{\sqrt{2b_2}}) - a_1 \sqrt{2b_2}\right\}
\nonumber \\
a_1 & = & (1 + b_1) - i\xi , \;\;\; b_1  =  \half d_1 \mu^2 t_1^3 =  
\half d_1 \mu^2 [(\tau + \Dl t)^3 + \tau^3] \nonumber \\
\eeqr
The left plot in Fig. \ref{fig: echopulse} shows the relative echo amplitude as a function of the
diffusion coefficient $D_n$ for three values of $n$. In each case, only the single $D_n$ was non-zero.
For the same value of $D_n$, the amplitude decreases faster as $n$ increases. 
The right plot in this figure shows the form of the echo pulse with the $D_1, D_2$ model for a particular
choice of $D_1, D_2$ and other machine parameters are taken from the RHIC values. The red curve
shows the upper envelope of the pulse which is used to obtain the full width at half maximum. 
\bfig[h]
\centering
\includegraphics[scale=0.55]{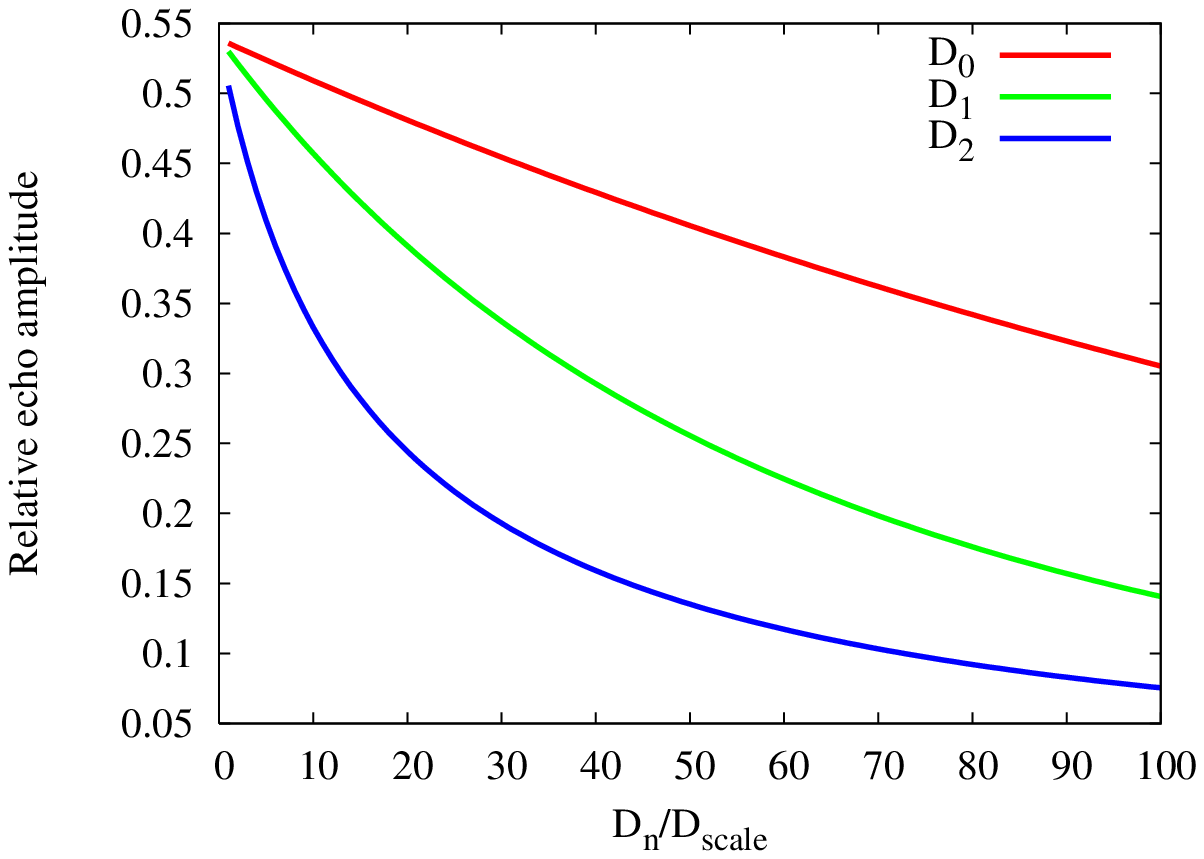}
\includegraphics[scale=0.55]{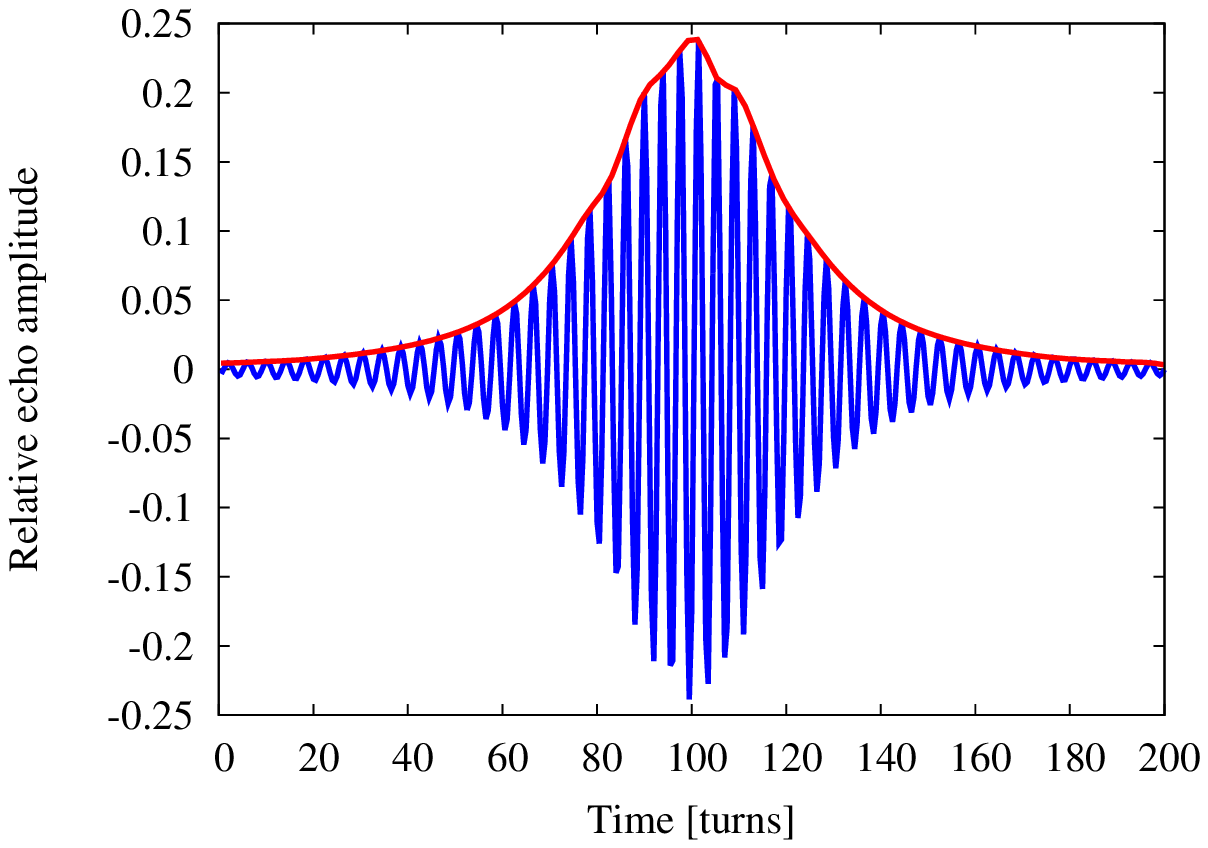}
\caption{Left: The echo amplitude as a function of the coefficients $D_0, D_1, D_2$ scaled by the
value $D_{scale}= 2.4\times 10^{-15}$m$^2$/s. 
Right: Form of the echo pulse with the $D_1, D_2$ model shown in blue. The red curve outlines the
upper envelope of the echo. Beam parameters in both plots were taken from 
Table I, except for $\mu=0.0077$. $D_1, D_2$ in the right plot were set to
the values $D_{1,sc}, D_{2,sc}$ respectively which are defined in Sec.
\ref{sec: opt_detune_delay}. }
\label{fig: echopulse}
\efig

\subsection{Optimum tune shift and delay time} \label{subsec: optima}

Analytical results for the optimum values of the tune shift and delay parameters that maximize the
echo amplitude can be obtained for model 1 with diffusion coefficients $(d_0, d_1)$. 
As a function of the time delay, this amplitude has a maximum at a delay $\tau = \tau_{opt}$, such that
the two coefficients can be related as 
\beq
d_1 = \frac{1 - 3 d_0 \mu_{fix}^2 \tau_{opt}^3}{\mu_{fix}^2 \tau_{opt}^3( 8 + 3 d_0 \mu_{fix}^2 \tau_{opt}^3)} 
\label{eq: d1_1}
\eeq
It is understood that $\mu$ is held fixed at $\mu_{fix}$ while finding the optimum delay 
$\tau_{opt}$. Defining  $c_{\tau} = \mu_{fix}^2 \tau_{opt}^3 $ and
substituting this into the equation for the relative amplitude, we have for the maximum amplitude obtained at the delay 
$\tau_{opt}$
\beq
\frac{\lan x\ran_{max}(\tau_{opt})}{\bt_K \theta} = \om_{rev}q \mu \tau_{opt}[\frac{8 + 3 d_0 c_{\tau}}{9}]^3
\exp[-d_0 c_{\tau}]
\label{eq: amp_d0_d1_tau}
\eeq
This equation can be solved for $d_0$ and subsequently $d_1$ can be found. Positivity of $d_1$ requires that the solution for $d_0$ obey $ 3 d_0 c_{\tau} \le 1$. 

Similarly, as a function of the tune shift, the amplitude has a maximum at $\mu = \mu_{opt}$ such that
\beq
d_1 = \frac{1 - 2 d_0 \mu_{opt}^2 \tau_{fix}^3}{\mu_{opt}^2 \tau_{fix}^3( 5 + 2 d_0 \mu_{opt}^2 \tau_{fix}^3)}
\label{eq: d1_2}
\eeq
Here $\tau$ is held fixed at $\tau_{fix}$ while finding the optimum in $\mu$. Defining
$ c_{\mu} = \mu_{opt}^2 \tau_{fix}^3 $
and again, substituting for $d_1$, we can write the maximum relative amplitude at $\mu_{opt}$ as
\beq
\frac{\lan x\ran_{max}(\mu_{opt})}{\bt_K \theta} = \om_{rev}q \mu_{opt} \tau_{fix}[\frac{5 + 2 d_0 c_{\mu}}{6}]^3
\exp[-d_0 c_{\mu}]
\label{eq: amp_d0_d1_mu}
\eeq
Here $d_1 \ge 0$ requires that the solution for $d_0$ obey $ 2 d_0 c_{\mu} \le 1$. 

If both $\mu_{opt}$ and $\tau_{opt}$ are measured, then the diffusion coefficient $d_0$ can be found
from equating the two expressions for $d_1$ which results in a quadratic equation for $d_0$
with the roots
\beq
d_0 = \frac{1}{12c_{\mu}c_{\tau}}\left[ 2c_{\mu}+3 c_{\tau} \pm \sqrt{\frac{(2c_{\mu}-3c_{\tau})(2c_{\mu}^2 + 67c_{\mu}c_{\tau} + 3c_{\tau}^2)}
{c_{\mu}- c_{\tau}}} \right] . \label{eq: d0_mum_taum}
\eeq
Once $d_0$ is determined, $d_1$ can be determined from either of Equations (\ref{eq: d1_1}) or 
(\ref{eq: d1_2}). Positivity of $d_1$ requires that the above solution obey
$d_0 \le 1/(2 c_{\mu})$ and $ d_0 \le 1/(3 c_{\tau}) $. 
This solution for both diffusion coefficients $d_0, d_1$ is obtained without necessarily 
using the value of echo amplitude except for recording where it has a maximum.  It uses the optimum tune shift and the optimum delay and could be 
useful when the BPM 
resolution is low. However this would require that all other beam conditions such as the 
dipole kick, quadrupole kick, bunch charge etc are kept exactly the same 
during both 
tune shift and delay scans. If this is not met, the solution given by 
Eq. (\ref{eq: d0_mum_taum}) cannot be used. 

For the $(d_0, d_2)$ or $(d_1, d_2)$ models discussed here,  the optimum values of the 
tune shift and delay parameters must be found numerically.

\subsection{Echo pulse width}

In addition to the amplitude, the echo can also be characterized by the echo pulse width, 
e.g the full width at half maximum (FWHM) can be chosen as a width measure. 

For the model $ D(J) = D_0 + D_1(J/J_0) $, 
the FWHM can be found analytically from Eq. (\ref{eq: amp_0_1}). We define a variable $D_{up}$
which depends on a upper limit to the pulse full width $(\Dl t)_{FW}^{up}$ 
and other parameters as follows 
\beq
D_{up} = (\frac{\eps}{\mu \om_{rev} \tau})^2 \frac{2}{(\Dl t)_{FW}^{up}}
\eeq
For example, with an upper limit to the pulse width of a 100 turns, we have 
$D_{up} = 2.6\times 10^{-12}$m$^2$/s.
For pulse widths $\Dl t_{FWHM} < (\Dl t)_{FW}^{up}$ such that $(D_0/D_{up}, D_1/D_{up}) \ll 1$,
we can keep terms to first order in 
$D_0/D_{up}, D_1/D_{up}$, and we find for the FWHM
\beq
\Dl t_{FWHM} = 2\sqrt{2^{2/3}-1}(\frac{\al}{\om_{rev}\mu}) + 3(\frac{\al \tau}{\om_{rev}})^2
\left[ \frac{2^{2/3}}{3}d_0 + \frac{d_1}{\al} \right], \;\;\;\;\;\;\;\;\;\; \;
\al  = 1 + \frac{1}{2}d_1 \mu^2 t_1^3
\label{eq: FWHM_0_1}
\eeq
As we see later, we have typically $(D_0/D_{up}, D_1/D_{up}) \approx 0.1$, so the above assumption
is satisfied for pulse widths up to a 100 turns or somewhat larger. We 
find that the
FWHM increases with increasing $D_1$ but very slowly with $D_0$ as seen in Fig. 
\ref{fig: fwhm_d0_d1_d2}.  When there is no diffusion, we have for the minimum
FWHM
\beq
\Dl t_{FWHM}^{min} =  \frac{2\sqrt{2^{2/3}-1}}{\om_{rev}\mu}
\label{eq: fwhm_min}
\eeq
In units of turns, this theoretical minimum FWHM depends only on the tune shift coefficient $\mu$. 
This value when compared with measured FWHM values can set limits on the tune shift parameter, as will be seen later. 

For the other models with either $(D_0, D_2)$ or $(D_1, D_2)$, the time dependent pulse shape and hence
the FWHM must be found numerically. From this pulse shape, the upper envelope is found numerically
as an interpolating function and the FWHM then calculated from this envelope function. 
Fig. \ref{fig: fwhm_d0_d1_d2} shows the dependence of the FWHM on the coefficients $D_0, D_1, D_2$
scaled by a parameter $D_{scale}= 2.4\times 10^{-15}$ m$^2$/s. The FWHM increases linearly
with both $D_0$ and $D_1$ but with $D_0$ increases by only 3\% over this range. The FWHM with $D_2$
increases the fastest and covers the range of values obtained from the RHIC data. 
\bfig
\centering
\includegraphics[scale=0.8]{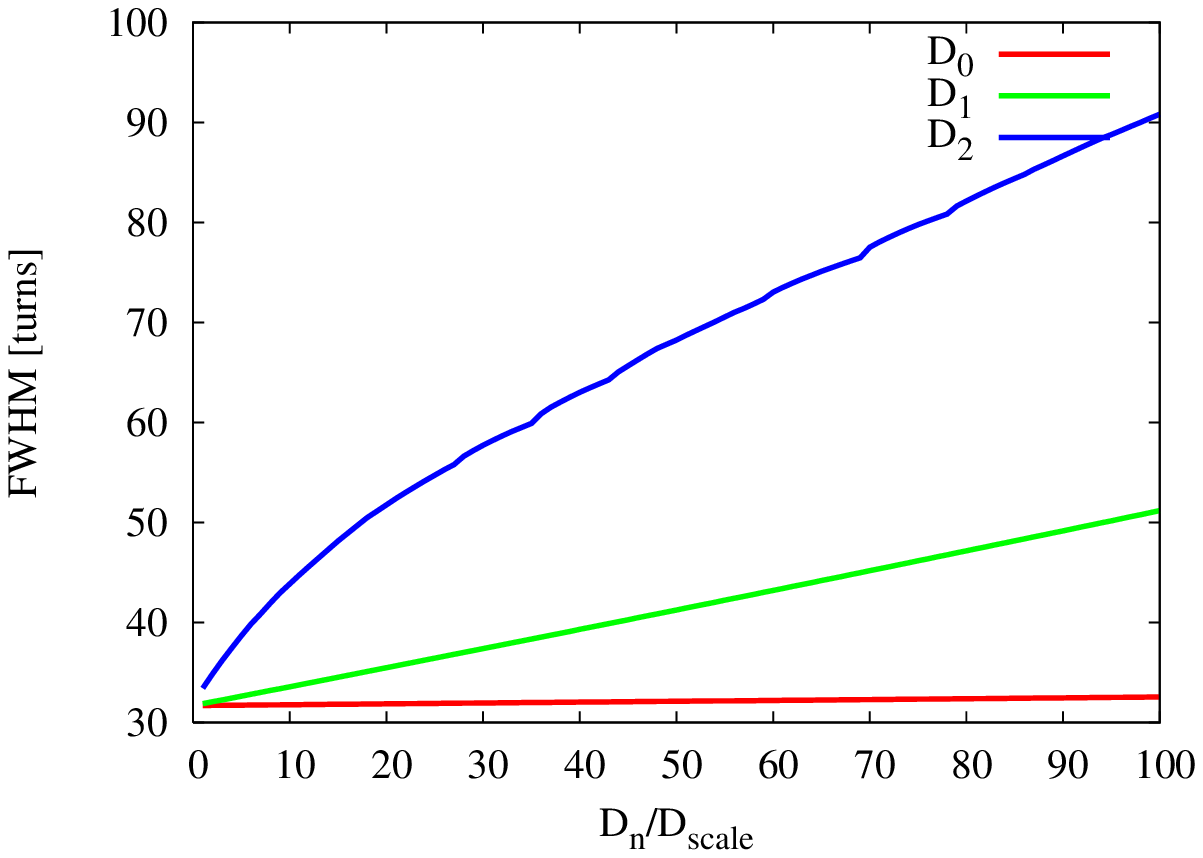}
\caption{FWHM as a function of the diffusion coefficients $D_0, D_1, D_2$ scaled by the 
value $D_{scale}$. Each curve shows the impact of the single coefficient with the others 
set to zero. 
The FWHM is calculated analytically from Eq. (\ref{eq: FWHM_0_1}) for $D_0, D_1$ and 
numerically for $D_2$. Parameters were taken from Table I, except for
$\mu=0.0077$.} 
\label{fig: fwhm_d0_d1_d2}
\efig

\section{Analysis of RHIC data with Au ions}
We briefly discuss the experimental procedure here, more details can be found in 
\cite{Fischer_PAC05}. The echo experiments were first done with Au ions, later with 
Cu ions and also with protons, all at injection energy.  A special purpose quadrupole 
kicker was used with a rise time of 12.8$\mu$s,  about one revolution time in RHIC. 
The nonlinear tune shift was provided by a set of octupoles which are normally set to zero at injection, in order to observe the echoes.
 The initial dipole kick was delivered only in the horizontal plane by injection under a varying angle. Echoes were generated with different conditions including
variable dipole and quadrupole kicks, beam intensities, tunes,  different delays between the
dipole kick and 
the quadrupole kick and different octupole strengths. The emittance delivered to RHIC for 
each species 
was nearly constant.   While echoes were observed with each species, the most consistent
echoes were obtained with the Au ions and we will consider only those results in this article.  Table \ref{table: RHIC} shows some of the relevant parameters for the Au ions 
\cite{Fischer_PAC05}. 
\begin{table}
\bec
\btable{|c|c|} \hline
Parameter &  Nominal Value \\ \hline
Beam relativistic $\gm$ &  10.52 \\
Revolution time $T_{rev}$ & 12.8 $\mu$s \\
Initial emittance $\eps_0$, un-normalized & 1.6$\times 10^{-7}$ m \\
Delay $\tau$ & 450 turns \\
Initial tune shift parameter $\mu_0$ & 0.0014  \\
Quadruple strength $q$ & 0.025 \\
Quadrupole rise time &  12.8 $\mu$s  \\
\hline
\etable
\eec
\caption{Relevant RHIC parameters for the echo experiments with Au ions.}
\label{table: RHIC}
\end{table}

\subsection{Emittance growth and rescaling tune shift}

In evaluating the tune shift parameter $\mu$ for calculating echo amplitudes,  it is important to use the emittance following the dipole kick. The rms emittance is given by 
\beqr
\eps & = & \frac{1}{\bt}[ \lan x^2 \ran \lan p^2 \ran - (\lan x p \ran)^2 ]^{1/2} \nonumber \\
 & = & 2 [\lan J \cos^2\phi \ran \lan J \sin^2 \phi \ran - \lan J\sin\phi\cos\phi \ran ^ 2]^{1/2}
\eeqr
The ensemble averages are calculated using the distribution function at 
time $t$ after the dipole kick which can be written in the notation of 
\cite{Chao} as
\beq
\psi_2(J,\phi, t) = \psi_0(J+\theta\sqrt{2\bt J}\sin(\phi - \om(J)t) + \half\bt_K \theta^2)
\eeq
and the averages are found from e.g. $\lan J\cos^2\phi \ran = \int dJ d\phi J \cos^2\phi \psi_2(J,\phi,t)$ etc. It can be shown this leads to an rms emittance given by 
\beqr
\eps(t) & = &  [(J_0 + \half \bt_K \theta^2)^2 - A_2(t)^2 ]^{1/2} \\
A_2(t) & = & \frac{\bt_K \theta^2}{2(1 + \Theta_2^2)^{3/2}}\exp[-\frac{\bt_K\theta^2}{2J_0}
\frac{\Theta_2^2}{1 + \Theta_2^2}], \;\;\;\;\;\;\;\;\; \Theta_2 = 2 \om' J_0 t \nonumber
\eeqr
At times $t \gg \tau_D$, the term $A_2 \rarw 0$ and we can approximate
\beq
\eps = J_0 + \half\bt_K \theta^2 = \eps_0 [ 1 + \half(\frac{\Dl x}{\sg_0})^2]
\label{eq: emit_kicked}
\eeq
where $\eps_0 = J_0$ is the initial emittance, $\Dl x= \sqrt{\bt_K \bt} \theta$ is the change in 
beam position at the BPM and $\sg_0= \sqrt{\bt\eps_0}$ is the initial beam size at the BPM. 
The last
expression in Eq. (\ref{eq: emit_kicked}) has the same form as in \cite{Edwards_Syphers}. 
Thus a kick to a 3$\sg$ amplitude results in an emittance which is 5.5 times larger than the
initial emittance. We will take this as an average estimate for the emittance following the dipole
kick. By definition, the tune shift parameter $\mu$ increases linearly with emittance and hence $\mu$
increases from its nominal value of 0.0014 to 0.0077 following the dipole kick. Without this rescaling,
the model cannot agree with the experimental results, as seen in the earlier analysis 
\cite{Fischer_PAC05, Sorge}.

\subsection{Diffusion coefficients from optimum tune shift and delay}
\label{sec: opt_detune_delay}

The theory predicts that the maximum echo amplitude, which occurs close 
to the time 2$\tau$ after the dipole kick, grows indefinitely with the 
product $\mu \tau$ in the absence of diffusion. In the presence of
any diffusion, the echo amplitude grows more slowly, reaches a maximum
and then decreases as either $\mu$ or $\tau$ is increased. In each case,
the irreversible particle motion caused by the presence of diffusion
reduces the amplitude of the recohering signal at the time of the echo.
Here we will apply the formulas developed in Section \ref{subsec: optima} 
to extract diffusion coefficients from measurements of the optimal
tune shift and optimal delay. 

We discuss first the analysis of the nonlinear tune shift scan done on March 11, 2004. During this scan, the quadrupole kick and delay between the dipole kick and quadrupole kick were kept constant. Octupole strengths were set to values $K_3=(1.5, 2, 2.5, 5, 6, 7, 8, 9, 10)$~m$^{-3}$. The nominal value was
$K_3= 7$~m$^{-3}$ corresponding to a nominal tune shift parameter $\mu_0= 0.0014$ before the
dipole kick. 
Echoes were observed for all $K_3 \ge 2.5$~m$^{-3}$. The largest echoes were observed at 
$K_3 = 5$~m$^{-3}$ which corresponds to a nominal tune shift parameter $\mu=0.001$ while the 
rescaled tune shift value is $\mu_{opt} = 0.0055$.

For the $D_0, D_1$ model, the starting solutions were obtained by solving 
Eqs. (\ref{eq: d1_2}) and (\ref{eq: amp_d0_d1_mu}). These yielded 
$d_0 = 2.245\times 10^{10}$~s$^{-3}$, $d_1=2.435\times 10^{10}$~s$^{-3}$, which lead to $D_0 = 1.08\times 10^{-13}$~m$^2$/s and $D_1 = 1.17\times 10^{-13}$~m$^2$/s. 
These found values for $(D_0, D_1)$ yield a maximum at $\mu_{opt}= 0.0055$ by design but the amplitude values decrease more slowly with $\mu$ than the data. To improve the fit with the data, a numerical 
fitting was done (using Mathematica \cite{Math}) to the data with the model shown in 
Eq. (\ref{eq: amp_0_1}). These yielded $D_0 = 1.62\times 10^{-13}$~m$^2$/s and 
$D_1 = 1.19\times 10^{-13}$~m$^2$/s and 
led to a better fit with all the data. These values for
$D_0, D_1$ were labeled as $D_{0,sc}, D_{1, sc}$ respectively and 
subsequent values were scaled by these values for convenience. With both the 
$(D_0, D_2)$ and the $(D_1, D_2)$ models, a least square minimization was done to fit 
the data against the respective models for the amplitude. The fit for $D_2$ from the
$(D_0, D_2)$ model was similarly labeled as $D_{2,sc}$. 
The resulting fits and the data are shown in Fig.\ref{fig: muscan_d0_d1_d2}. The values of the 
coefficients are shown in Table \ref{table: Au_D0_D1_D2_scans}.
\bfig
\centering
\includegraphics[scale=0.8]{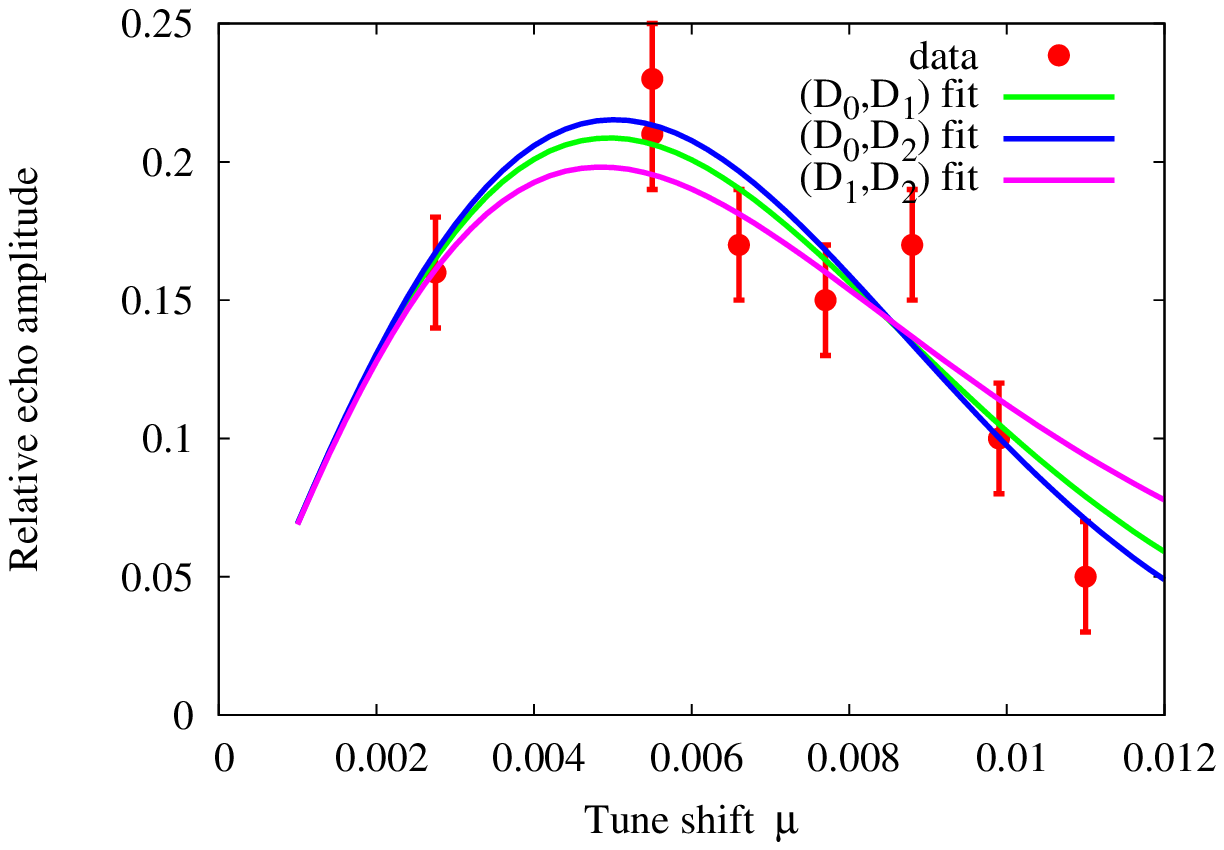}
\caption{Comparison of the echo amplitude vs tune shift strength scan. The data shown in red 
with error bars while the fits shown are with the three models for
the diffusion coefficients discussed in the text.}
\label{fig: muscan_d0_d1_d2}
\efig
Relative to the previous comparison of theory with experimental data 
cf. Fig. 4 in \cite{Fischer_PAC05}, these fits show significant improvement. Of the three models, the best fit with the lowest chi squared is seen with the $(D_0, D_2)$ model with the next best being the $(D_1, D_2)$ model. However the models are fairly close and no model can be ruled out based on this data.

On a later day (March 17, 2004), the delay $\tau$ between the dipole kick and the 
quadrupole kick was varied with values (450, 500, 550, 600, 900) turns. Echoes were only 
observed at the first three values of the delay. In all six echoes were observed with the 
largest amplitudes at 450 turns. The quadrupole kick strength, the octupole strengths 
and the tunes were kept constant. We will use this limited data set to obtain the diffusion 
coefficients from the delay scan. 

For the $(D_0, D_1)$ model, we start by solving Eq. (\ref{eq: d1_1}) and 
Eq. (\ref{eq: amp_d0_d1_tau}) for the coefficients from the echo amplitude and the value of the optimum delay $\tau_{opt}$. Again, better fits to the data are obtained by a least square minimization which is also the procedure for the other two models. Table \ref{table: Au_D0_D1_D2_scans} shows the best
fit values with this delay scan. Compared to the values from the tune shift scan, the coefficients for the same model are within a factor of two from this delay scan. Some of the variation in the values between the
scans can be due to different beam conditions on the two days such as bunch intensities and machine 
tunes.
However the uncertainties associated with these values are large since there were too few data points. 
Fig. \ref{fig: tauscan_d0_d1_d2} shows the comparison of the fitted models with the data. 
\bfig
\centering
\includegraphics[scale=0.8]{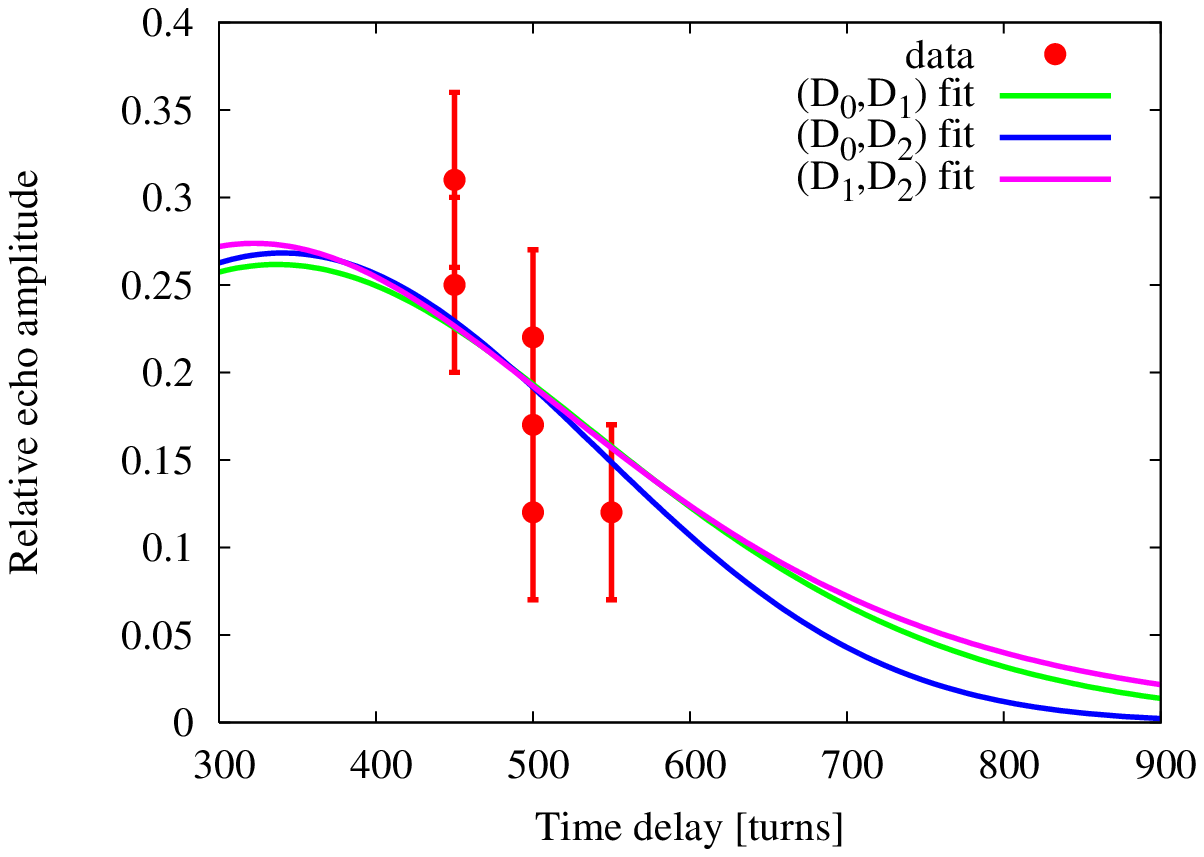}
\caption{Comparison of the echo amplitude vs time delay between the dipole and quadrupole kicks.
The data shown with error bars while the fits are shown with the three models discussed in the text.}
\label{fig: tauscan_d0_d1_d2}
\efig
Again all three models show similar goodness of fits with the best fit (minimum chi squared) obtained with the $(D_0, D_2)$ model but all chi squared values are close. 
All models show that the relative echo amplitude reaches a maximum at around 390 turns which is less than the minimum delay of 450 turns used in the experiment. 
\begin{table}
\bec
\btable{|c|c|c|} \hline
Model & Tune Shift scan & Delay scan \\ \hline
$D_0$ / $D_1$  & 1.6 / 1.3 &  0.65 / 1.3 \\
$D_0$ / $D_2$  & 1.9 / 0.025 & 3.7 / 0.015 \\
$D_1$ / $D_2$  & 2.3 / 0.025 & 1.9 / 0.013 \\
\hline
\etable
\eec
\caption{Comparison of the diffusion coefficients from the tune shift and delay scans. All diffusion
coefficients are in units of $10^{-13}$~m$^2$/s. }
\label{table: Au_D0_D1_D2_scans} 
\end{table}

\subsection{Diffusion coefficients from the echo amplitude and the FWHM }
\label{subsec: amp_fwhm}

The above analysis has shown that all three models are viable candidates in describing the 
data dependence on either the tune shift or the delay. We now use turn by turn (TBT)  data to 
fit both the echo amplitude and the echo pulse width with each model. Ten such data sets 
could be retrieved from the 2004 measurements. In this TBT set, the 
initial dipole kick and bunch charge varied but the 
other parameters including the quadrupole kick strength, tunes, delay and
octupole strengths were kept 
constant. Figures \ref{fig: echopulse_1_fwhm} and  \ref{fig: echopulse_2_fwhm} show two examples from this 
set, one with a clean echo pulse and the other where the beam centroid takes a longer time to decohere after the initial kick and the echo pulse is also much wider. Some of the more distorted signals could be due to oscillations from off-axis injection and could partly be due to a fourth order 
resonance and slightly higher bunch  charge. 
\bfig
\centering
\includegraphics[scale=0.5]{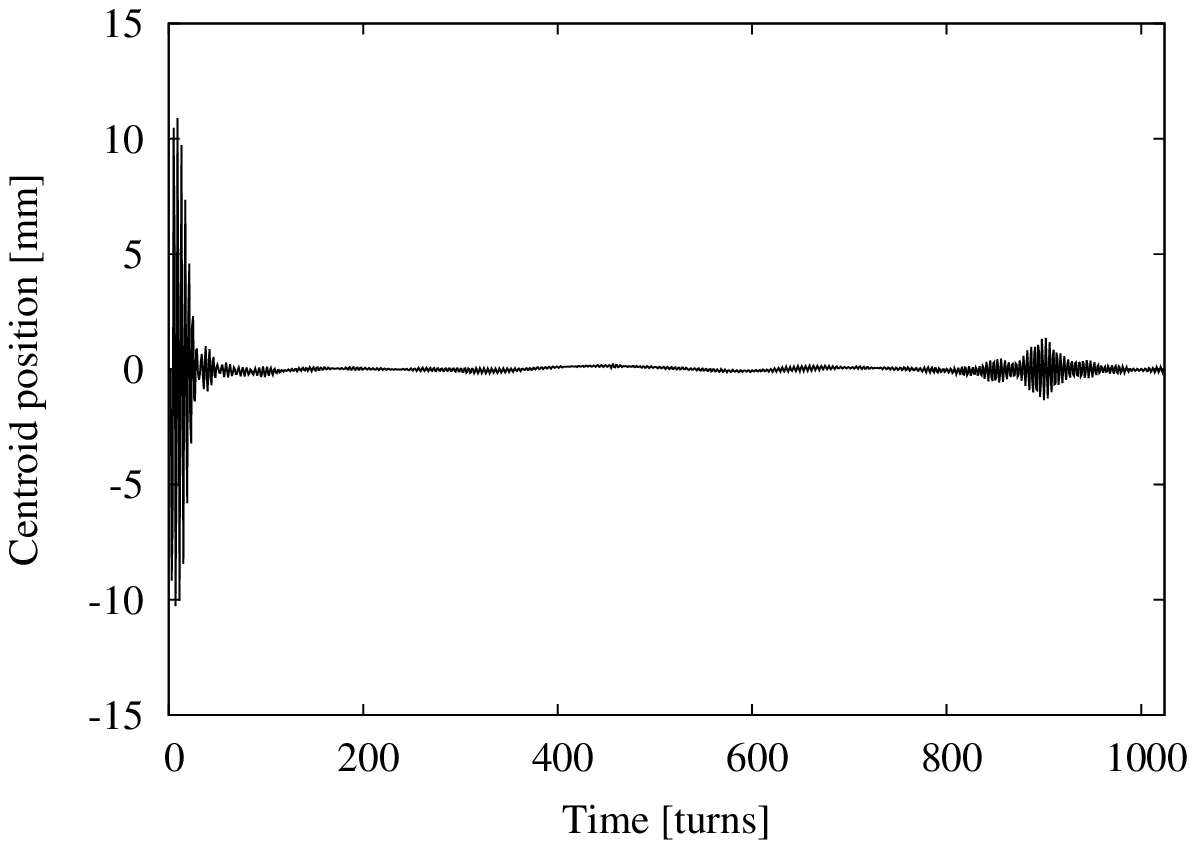}
\includegraphics[scale=0.5]{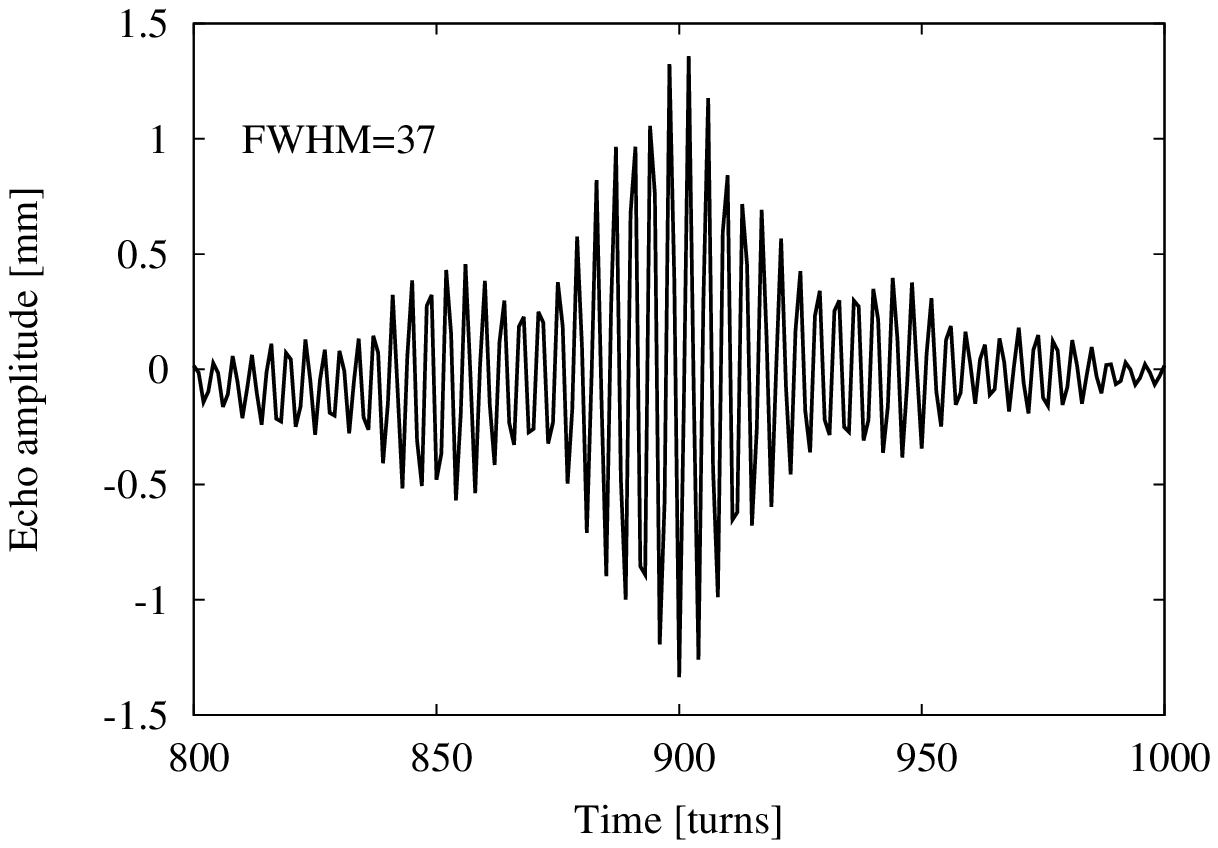}
\caption{The entire centroid position turn by turn (left) and the echo pulse isolated (right) for
the data with the shortest FWHM. Here the centroid decoheres cleanly after the dipole kick.
The quadrupole kick was applied at turn 450.}
\label{fig: echopulse_1_fwhm}
\includegraphics[scale=0.5]{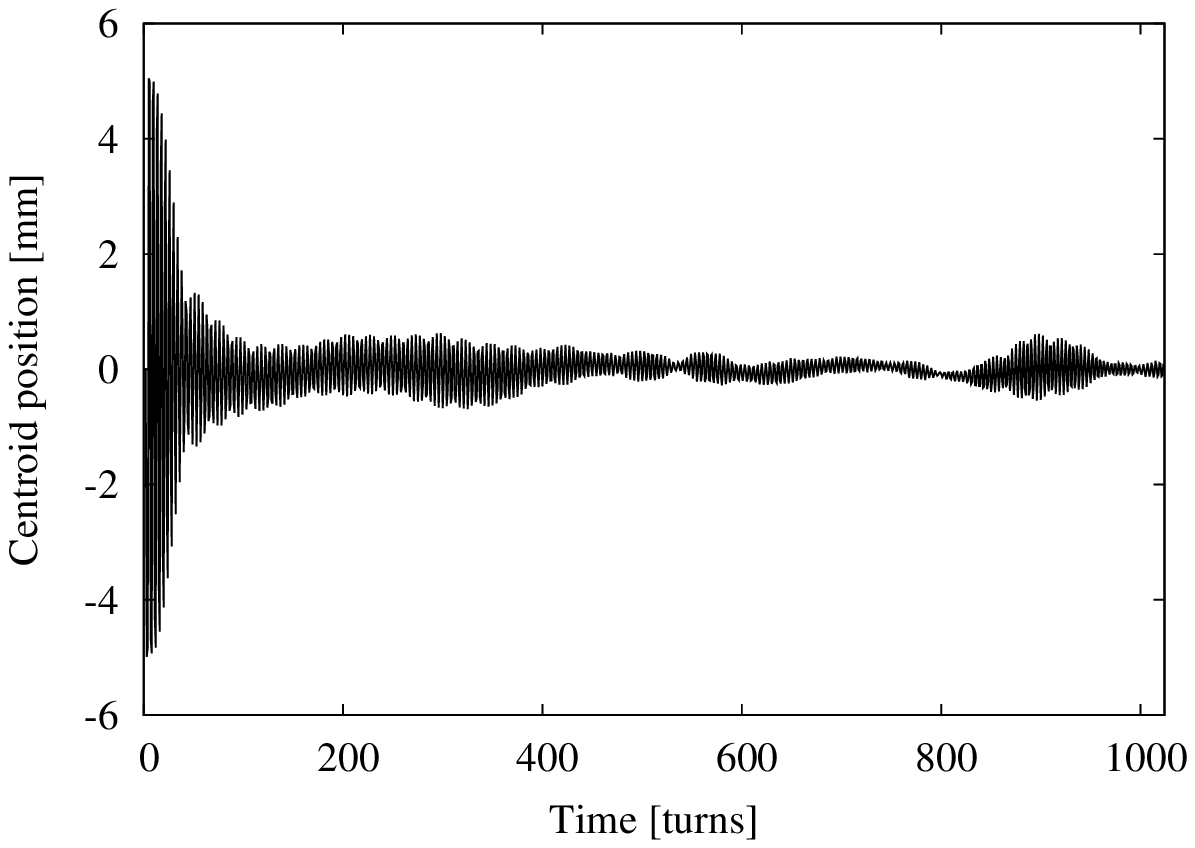}
\includegraphics[scale=0.5]{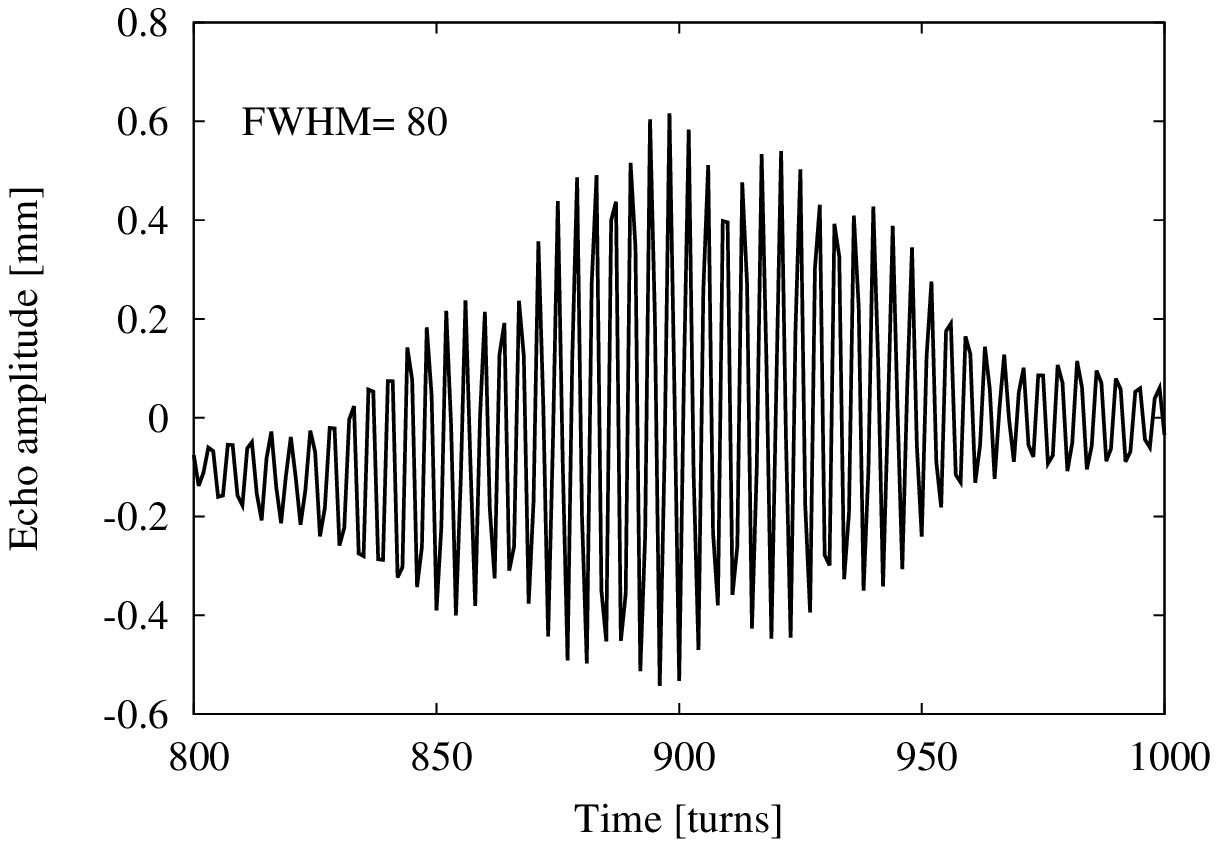}
\caption{The entire centroid position turn by turn (left) and the echo pulse isolated (right) for
the data with the largest FWHM. Notice the much larger and longer ringing of the centroid after the
dipole kick. The quadrupole kick was applied again at turn 450. }
\label{fig: echopulse_2_fwhm}
\efig
For each data set, an interpolating function was found to fit the upper envelope of the echo pulse and the  FWHM was extracted from this interpolating function.
Using the value of the rescaled tune shift parameter $\mu = 0.077$, the minimum theoretical value of
the FWHM without diffusion, using Eq. (\ref{eq: fwhm_min}), is 32 turns. This is consistent 
with the minimum FWHM with diffusion from 
the data set which is 37 turns. The bare tune shift parameter of $\mu_0= 0.0014$ would have 
predicted a minimum FWHM of 160 turns, much larger than any FWHM value measured.

Fig. \ref{fig: fwhm_intensity} shows the FWHM plotted as  a function of the
number of particles per bunch.
This figure shows that the FWHM fell into three distinct clusters because the bunch charge varied  around three values.  Except
for the two outlier points labeled as (1, 2), all other points show that the FWHM increases with
charge. These other points are fit to a power law curve 
\beq
FWHM(N) = \Dl t_{FWHM}^{min} + a N^p 
\eeq
where $\Dl t_{FWHM}^{min}$ is the minimum FWHM from Eq. (\ref{eq: fwhm_min}), $N$ is the number
of particles per bunch and $(a,p)$ are the fit parameters. 
The fit shows that the exponent is $p= 2.002$, so the FWHM increases quadratically with the
charge. 
\bfig
\centering
\includegraphics[scale=0.8]{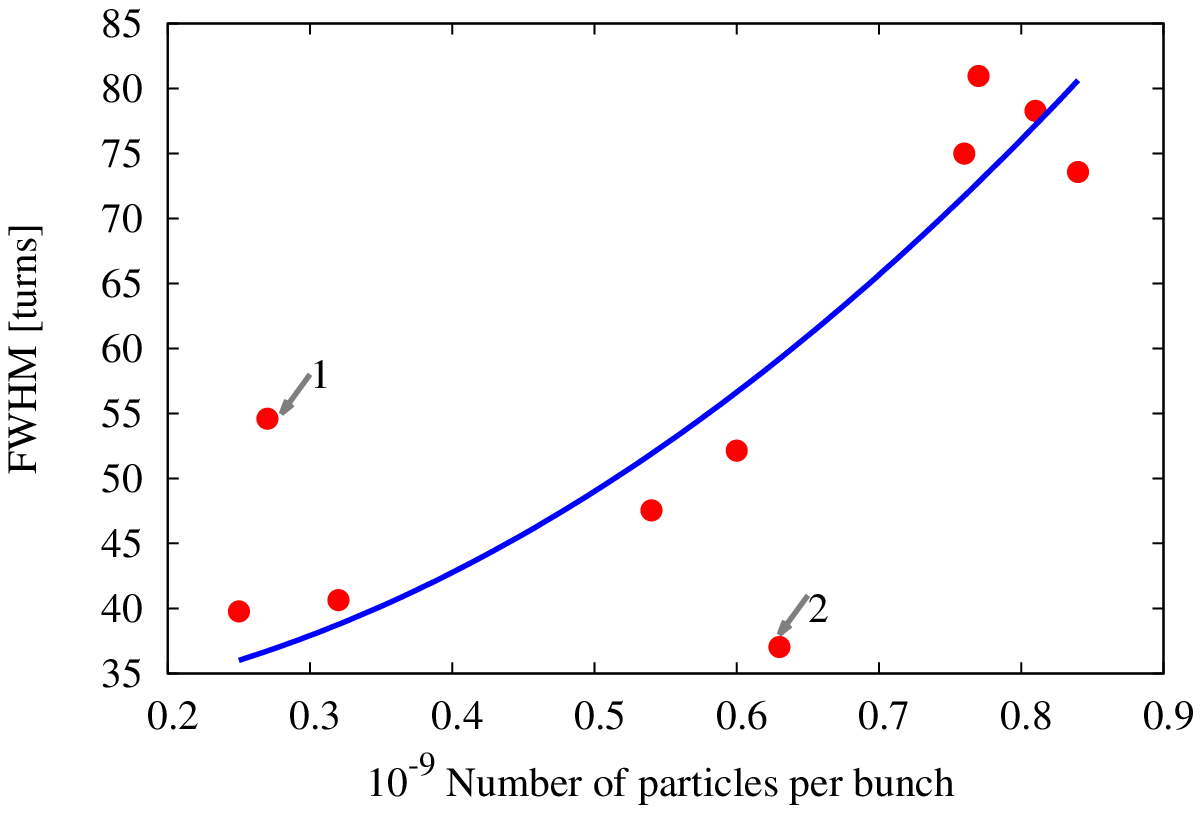}
\caption{FWHM vs the number of particles per bunch. Except for the points labeled 1 and
2, all the other data points show an increasing FWHM with bunch charge. The blue curve shows a 
quadratic fit through these points. }
\label{fig: fwhm_intensity}
\efig
Since the tune shift, delay,and tune were kept constant during these measurements, the outlier points 
show that the FWHM values may depend on other parameters, such as the initial dipole kick amplitude.

We now solve for two diffusion coefficients using the relative echo amplitude and the FWHM. 
For the $(D_0, D_1)$ model, the FWHM can be found 
analytically, as shown in Eq. (\ref{eq: FWHM_0_1}). The $d_0$ coefficient can be written as a function of
the echo amplitude and $d_1$ using the echo amplitude equation 
Eq.(\ref{eq: amp_0_1}) as
\beq
d_0 = - \frac{1}{\mu^2 \tau^3} \ln[\frac{\lan x\ran^{amp}_{rel}}{2\pi q \mu N_{delay}}
(1+\mu^2\tau^3 d_1)^{3/2}]
\label{eq: d0_from_d1}
\eeq
where $\lan x \ran^{amp}_{rel} = \lan x \ran^{amp}/(\bt_K \theta)$ is the relative echo
amplitude in terms of the dipole kick and $N_{delay} = \tau/T_{rev}$ is the delay in units of 
turns. The positivity of $d_0$ implies an upper limit to $d_1$ as
\beq
d_1^{max} = \frac{1}{\mu^2 \tau^3}[ \left(\frac{2\pi q \mu N_{delay}}{\lan x\ran^{amp}_{rel}}\right)^{1/3}-1]
\eeq
The value of $d_1$ can be found by numerically solving the equation Eq. (\ref{eq: FWHM_0_1})
for the FWHM  with
$d_0$ substituted from Eq. (\ref{eq: d0_from_d1}). We find that this $(D_0, D_1)$ model yields
positive $d_0$ coefficients in only four of the ten cases. We conclude therefore that the $D_0, D_1$ 
model is not well suited for this data. 

With the $(D_0, D_2)$ model,the $d_0$ coefficient can again be found analytically as a function of
the echo amplitude and $d_1$ using
\beq
d_0 = - \frac{1}{\mu^2 \tau^3} \ln \left[\frac{\lan x\ran^{amp}_{rel}}{\pi q \mu N_{delay}}
\frac{1}{{\rm Im}[e^{i\Phi_0(T_{rev})}H_{02}(T_{rev})]} \right]
\label{eq: d0_from_d2}
\eeq
where $H_{02}$ is defined in Eq. (\ref{eq: amp_0_2}). 
We find again that no solutions with positive $D_0$ can be found in all cases with 
FWHM $>$ 70 turns. Even in other cases where the solutions can be found, the values of 
$D_2$ are significantly larger than the values found in the previous sections, hence appear 
to be in a disconnected region of the parameter space. Since $D_0$ has little impact on the 
FWHM (see Fig.  \ref{fig: fwhm_d0_d1_d2}), in both the $(D_0, D_1)$ and $(D_0, D_2)$ models,
large values of the FWHM can make $D_1$ or $D_2$ large which then require a negative $D_0$ to satisfy the amplitude condition. Thus fitting the models to both the amplitude and FWHM rules out the models with
 $D_0$. 

In the case of the $D_1, D_2$ model, neither coefficient can be found analytically from
 the amplitude equation. Instead the amplitude and the FWHM equations must be solved
numerically. Figure \ref{fig: amp_fwhm_3D} shows the forms of the function 
${\rm ampl}(d_1, d_2)$ and ${\rm fwhm}(d_1, d_2)$. Also shown are the intersections of these
surfaces with the plane of constant amplitude or FWHM value respectively. In each case, the 
intersection 
of the surface with the plane determines a curve of solutions for that equation. The 
intersection of the two curves in the $d_1, d_2$ plane would determine the required solution
for given values of the amplitude and FWHM. In this figure the values of $d_1, d_2$ are 
scaled by $d_{1,sc}, d_{2,sc}$ which are obtained from $D_{1,sc}, D_{2,sc}$ using Eq. (\ref{eq: dn_Dn}). These plots demonstrate that for the range of measured values of the
echo amplitude and the FWHM, solutions for the diffusion coefficients exist in the range 
$0 \le (d_1/d_{1,sc}, d_2/d_{2,sc}) \le 8$. 
\bfig
\centering
\includegraphics[scale=0.55]{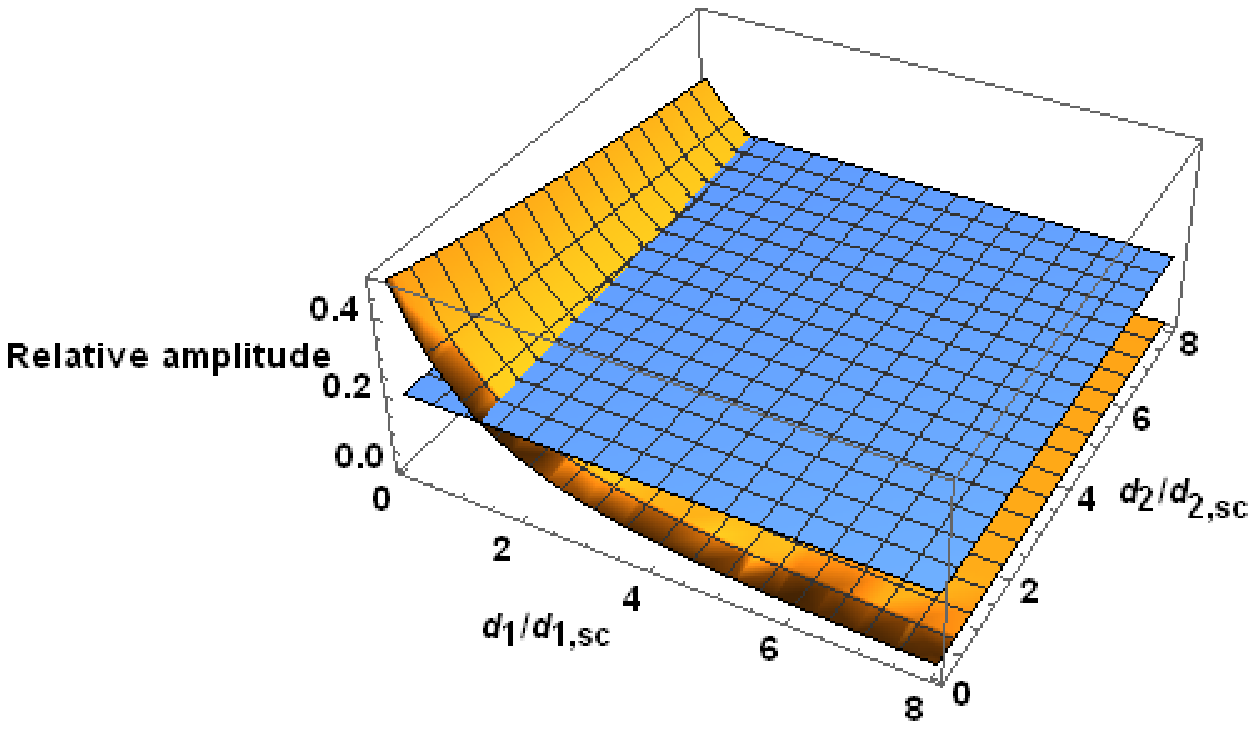}
\includegraphics[scale=0.55]{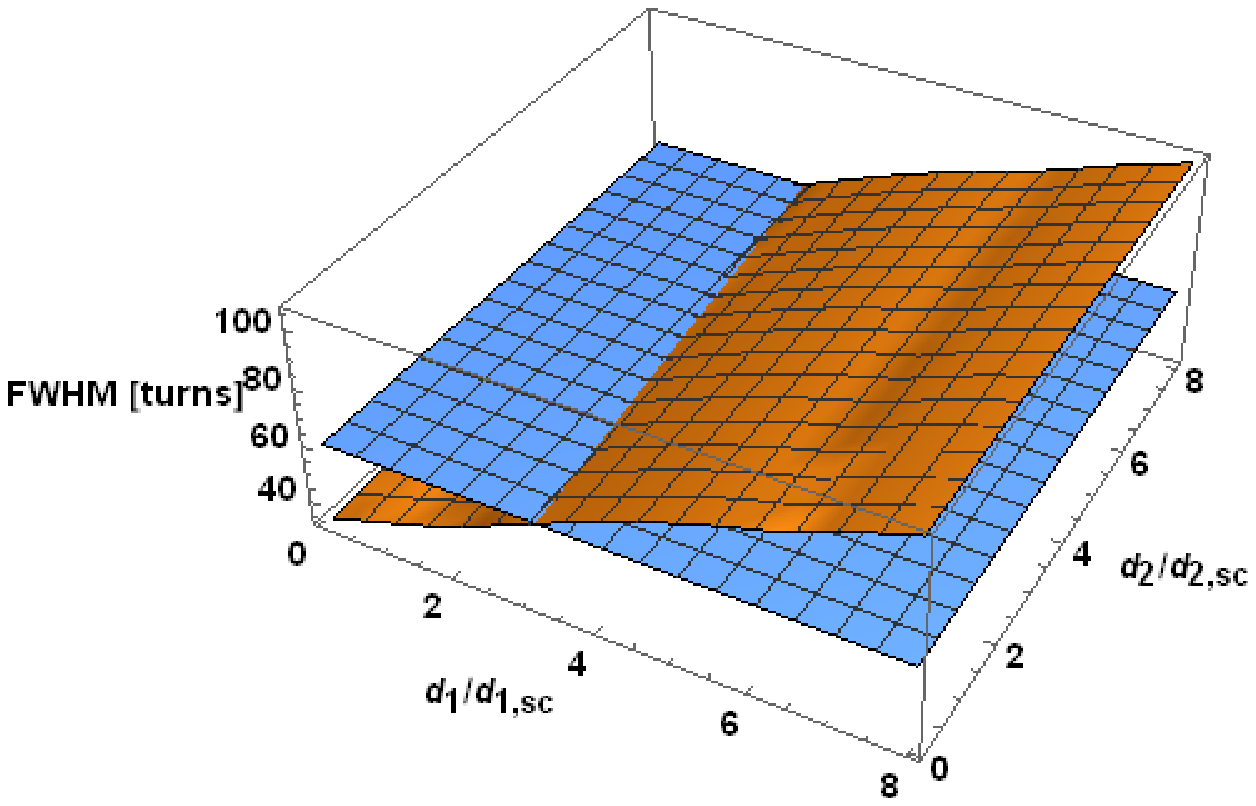}
\caption{Left: Relative echo amplitude (in brown) as a function of the scaled diffusion 
coefficients
$d_1/d_{1,sc}, d_2/d_{2,sc}$ intersected by a plane (in blue) of a particular relative 
amplitude value, here chosen to be 0.2. The intersection defines the family of solutions for
$(d_1, d_2)$ at this amplitude. 
Right: The FWHM (in brown) as a function of the same scaled variables and the plane 
(in blue) at a constant FWHM, here chosen to be 60 turns. Again, the intersection defines 
the family of solutions for the FWHM equation.}
\label{fig: amp_fwhm_3D}
\efig

It turns out to be easier to do a least squared minimization to find the solution. Here we define the $\chi^2$ function as
\beq
\chi^2 = (\frac{{\rm ampl}(d_1,d_2) - {\rm ampl_{data}}}{\sg_{ampl}})^2 +
 (\frac{{\rm fwhm}(d_1,d_2)- {\rm fwhm_{data}}}{\sg_{fwhm}})^2 
\eeq
where  ampl$(d_1, d_2)$ and fwhm $(d_1, d_2)$ are the amplitude function (from 
Eq. (\ref{eq: amp_d1_d2}) ) and
 the FWHM function defined numerically and $\sg_{ampl} = 0.05$ and $\sg_{fwhm}=2$ are 
the estimated uncertainties in the two data variables. This least squares method turns
 out to be efficient and leads to positive solutions for $d_1, d_2$ in all cases.  
Table \ref{table: D1_D2_amp_fwhm} shows the values of the diffusion coefficients in these cases. We observe that these values are close to the values of $D_1$ found from the
optimal tune shift and delay measurements shown in Table \ref{table: Au_D0_D1_D2_scans}. 
The $D_2$ values differ by an order of magnitude in the two tables but considering that
the delay and tune shift scan methods for the amplitude are less sensitive to $D_2$ and
also from the larger number of data points in the FWHM analysis, we expect the values in
Table \ref{table: D1_D2_amp_fwhm} to be more accurate. 
In most cases, the $D_1$ coefficient is an order of magnitude greater than $D_2$. 
The single exception (row 2 of this table) corresponds to the outlier
point labeled 1 in Fig. \ref{fig: fwhm_intensity}. As a function of charge, $D_1$ increases while 
$D_2$ appears to be independent of the charge. 
\begin{table}
\bec
\btable{|c|c|c|c|c|} \hline
Particles per bunch  & Rel. ampl. & FWHM & $D_1$ & $D_2$ \\ 
\mbox{} [$10^{9}$]  & [ ] & [turns] & [$10^{-13}$~m$^2$/s] & 
[$10^{-13}$~m$^2$/s] \\ \hline
0.25  &     0.245 &     39.8   &    1.28      &     0.0030   \\
0.27  &    0.225  &     54.6   &    0.13    &     0.51      \\
0.32  &    0.160  &    40.6    &     1.49     &     0.32   \\
0.54  &    0.127  &      47.5   &     2.00   &      0.28  \\
0.6    &    0.142   &      52.1   &    1.98   &       0.21    \\
0.63  &    0.125   &      37.0    &    1.98  &      0.30    \\
0.76  &   0.114    &      75.0    &    2.53  &     0.24   \\
0.77  &    0.122   &     81.0     &    2.18  &    0.24  \\
0.81  &    0.110   &     78.3     &     2.53  &     0.24    \\
0.84  &    0.0998  &     73.6   &     2.53    &       0.24  \\
\hline
\etable
\eec
\caption{Diffusion coefficients $(D_1, D_2)$ found using the amplitude and the FWHM values from the 
turn by turn data.}
\label{table: D1_D2_amp_fwhm} 
\end{table}

%\clearpage

\subsection{Diffusion dependence on bunch charge} \label{sec: intensity}

We focus now on the $(D_1, D_2)$ model which is the only one of those 
studied that can describe both the amplitude and pulse width of the echo. During the measurements on March 17, 2004 an intensity scan was done 
with all other parameters kept constant. While the turn by turn data from that scan is not easily
accessible, the echo amplitudes are available with 27 data points. This data can be used to
measure the diffusion coefficients as a function of bunch charge. 

Both $(D_1, D_2)$ coefficients can be found by a least square minimization of the fit to the amplitude.
This process allows a determination of $(D_1, D_2)$ as a function of charge,  The left plot in 
Fig. \ref{fig: Au_fit_intensity} shows the $D_1$ values found and a linear fit to the values. This confirms the behavior seen in the previous section but now with a larger data set. Similarly as earlier, the $D_2$ values are nearly independent of the charge. We can parameterize the echo amplitude's 
dependence on bunch charge via these fits for $D_1, D_2$ and the amplitude equation 
(\ref{eq: amp_d1_d2}). The linear fit yields $d_1/d_{1,sc} = 0.42 + 2.78 N$ where $N$ is the
number of particles per bunch in units of 10$^9$ while for $d_2$ we take the mean value over this set, 
$d_2/d_{2,sc}= 6.24$
The right plot in Fig. \ref{fig: Au_fit_intensity} shows the measured echo
amplitudes (in red) as a function of the number of particles per bunch and also the calculated 
amplitude (in blue) from these  fits for $(D_1, D_2)$. 
 The measured echo amplitude decreases with increasing charge, and this trend is well 
reproduced by the theoretical amplitude function. This is a consistency check and is to be
expected, since the
linear fit for $d_1$ and constant for $d_2$ were obtained from the data set. 
\bfig
\centering
\includegraphics[scale=0.55]{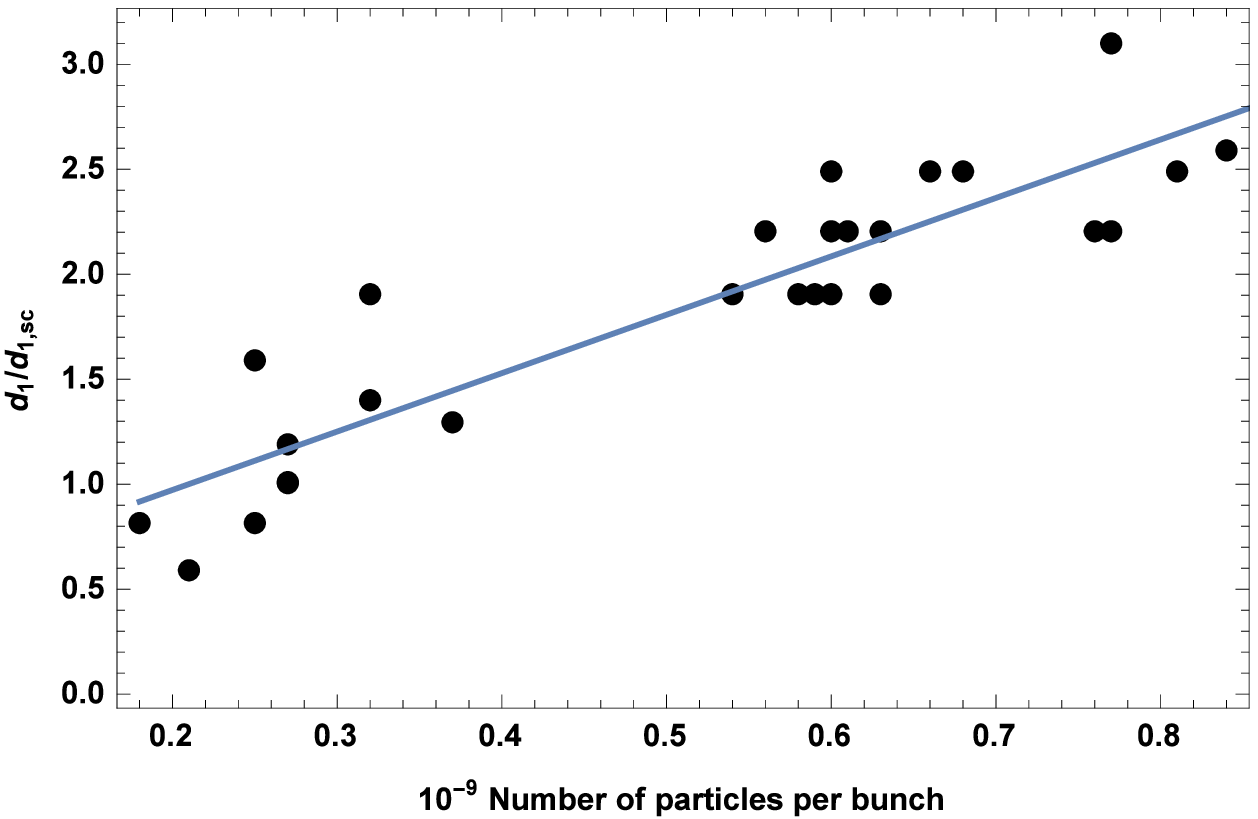}
\includegraphics[scale=0.55]{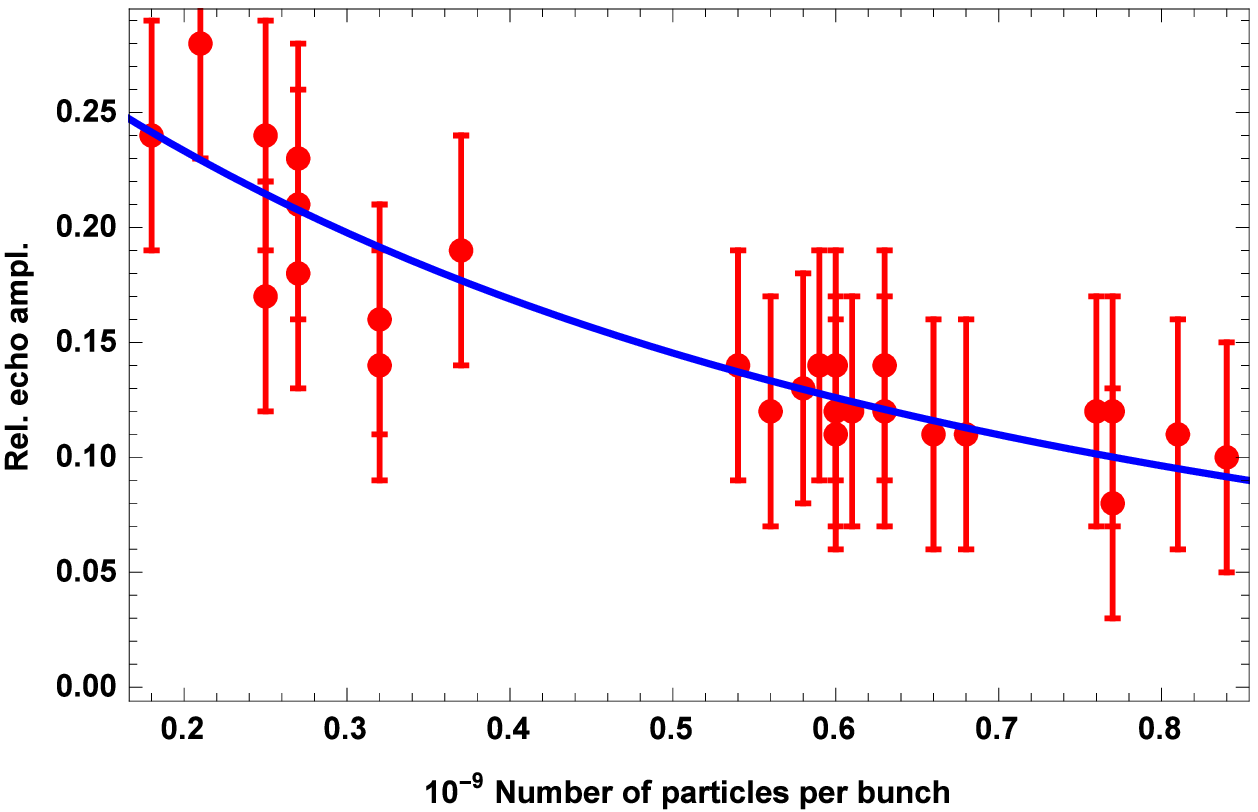}
\caption{Left: Calculated $d_1$ values as a function of the number of particles per bunch and the 
linear fit to the values. Right: 
Measured relative echo amplitude (red) at different intensities and compared with the
best fit curve (blue) with $(D_1,  D_2)$, with D1 from the linear fit in the left plot
and D2 independent of the bunch charge.}
\label{fig: Au_fit_intensity}
\efig
The comparison in Fig. \ref{fig: Au_fit_intensity} shows that we can parameterize the diffusion coefficients as 
\beq
D(J) = [a_{10} + a_{11}N](\frac{J}{J_0}) + a_{20}(\frac{J}{J_0})^2 
\eeq
where $a_{10}, a_{11}, a_{20}$ are functions of machine and beam parameters such as the nonlinearity, tunes, emittance etc. but independent of the bunch charge.

Space charge effects and intra-beam scattering (IBS) are the dominant source of particle 
diffusion for heavy ions such as Au in RHIC, at injection energy. The incoherent space 
charge  and IBS induced diffusion 
and emittance growth depends linearly on the charge and our analysis 
confirms that the leading diffusion coefficient $D_1$ increases linearly with
charge. The coefficient $D_2$ is likely to be determined by diffusion from single
particle  nonlinear dynamics processes. 

In the above analysis we have neglected the effect of wakefields on the 
echo formation. Their impact on the calculations above is not  likely to be 
significant. As seen in Figures \ref{fig: echopulse_1_fwhm} and 
\ref{fig: echopulse_2_fwhm} and generally true for the available
turn by turn data, the centroid response after the dipole kick is 
cleaner and the relative echo amplitude is larger 
with the larger amplitude kick. This would likely not be the case if
the effects of the transverse wake were significant. Instead, effects
due to injection oscillations and fourth order resonance which shows up at intermediate amplitudes
are the likely reason for the response seen in Fig. 
\ref{fig: echopulse_2_fwhm}.
In addition, the effect of wake fields would be visible in a change in the 
decoherence time with intensity. An analysis of the intensity scan data 
shows no correlation between the decoherence time and the bunch intensity.

\section{Mean escape time}
One useful time scale that can be extracted from the diffusion coefficients is the mean
escape time $t_{esc}$ associated with probabilistic processes \cite{Gardiner}. This time, also 
known as the 
mean first passage time, is the mean time taken (averaging over many realizations of the process)
for a particle to escape from a certain region defined by a boundary. 
It was shown in
\cite{Sen_Tev} that in the case that $D(J) = D_1(J/J_0)$, the time dependent 
density distribution solution $\psi(J,t)$ to the diffusion equation leads to a
beam lifetime $t_L$ which is close to the escape time $t_{esc}$ estimate. Defining 
$t_L = -N(t)/(dN/dt)$ where $N(t) = \int \psi(J,t) dJ$ is the particle number, it was shown that
\beq
t_L \approx 0.7 \frac{J_A J_0}{D_1}, \;\;\;  t_{esc} = \frac{J_A J_0}{D_1}
\eeq
where $J_A$ is the action at the absorbing boundary. 
We will assume that the mean escape time
is also a useful beam relevant time scale when $D(J) = D_1(J/J_0) + D_2(J/J_0)^2$.

The mean escape time from an action $J$ to an absorbing boundary at action $J_A$ is given by
\beqr
t_{esc}(J) &  = & \int_J^{J_A} dJ \frac{J}{D(J)} = \int_J^{J_A} dJ \frac{J}{D_1 (J/J_0) + D_2 (J/J_0)^2} \nonumber \\
 & = & \frac{J_0^2}{D_2}\ln[\frac{D_1 + D_2 (J_A/J_0)}{D_1 + D_2(J/J_0)}]
\eeqr
This is the mean escape time for particles initially at a single action 
$J$ to reach the aperture at action $J_A$ due to diffusion. 
A parameter describing the escape time for the beam
can be obtained by averaging this over the initial beam distribution $\psi_0(J)$, which yields
\beqr
\lan t_{esc} \ran  & = & \frac{J_0}{D_2}\int_0^{\infty}  dJ\; \exp[-\frac{J}{J_0}]
\ln[\frac{D_1 + D_2 (J_A/J_0)}{D_1 + D_2(J/J_0)}] \nonumber \\ 
 & = & \frac{J_0^2}{D_2}\left[ \ln(\frac{D_1}{D_2} + \frac{J_A}{J_0}) - \ln\frac{D_1}{D_2} - 
 e^{D_1/D_2}\Gamma(0, \frac{D_1}{D_2}) \right]  \equiv \frac{J_0^2}{D_2} A_F
\label{eq: escape}
\eeqr
where $\Gamma(0,z)$ is the incomplete Gamma function and we have assumed $D_2 \ne 0$.
The dimensionless amplifying factor $A_F$, defined by the terms in square brackets,
depends only the ratios $D_1/D_2, J_A/J_0$, Figure
\ref{fig: escape_D1_D2} shows the dependence of the dimensionless terms on $D_1/D_2$
for three values of $J_A/J_0$ corresponding to apertures at (6,10, 12)$\sg$
respectively. 
\bfig
\centering
\includegraphics[scale=0.75]{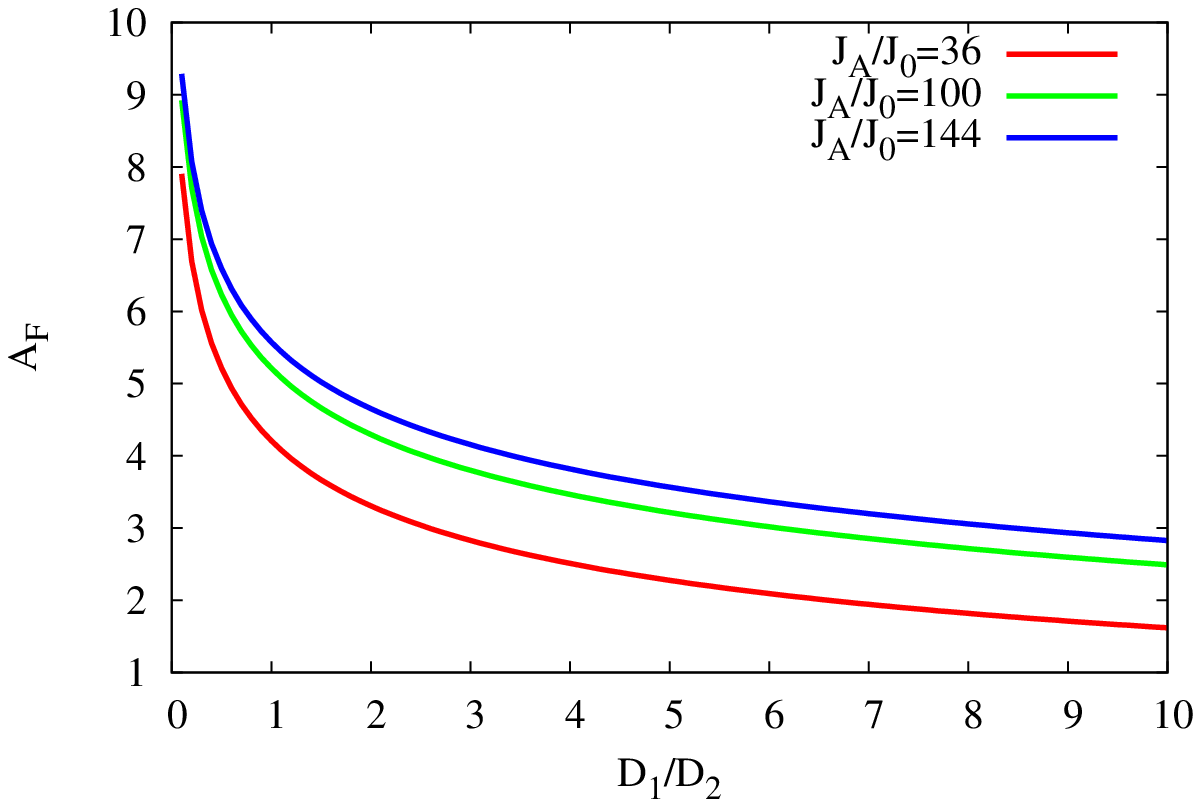}
\caption{Dependence of $A_F$, defined in Eq. (\ref{eq: escape}), on the ratio of 
diffusion
coefficients $D_1/D_2$ for three values of the ratio of the action at the absorbing
aperture $J_A$ to the initial emittance $J_0$.}
\label{fig: escape_D1_D2}
\efig
For $D_1/D_2 \simeq 10$, $A_F$ is of order unity. Hence the mean escape time is 
determined primarily by $J_0^2/D_2$. In the case that $D_2=0$, the time scale would
be determined by $J_0 J_A/D_1$.  With $J_0 = 1.6\times 10^{-7}$m, and taking
a representative value $D_2 = 0.25\times 10^{-13}$~m$^2$/s from Table 
\ref{table: D1_D2_amp_fwhm}, we have $\lan t_{esc} \ran \approx 1$s. 
While this time is extremely short, it corresponds to the lifetime of a beam at large amplitudes
and not to a beam circulating on the nominal closed orbit. Observations in RHIC did show
that lifetimes of kicked beams were significantly smaller compared to that for beams not kicked.
However the early losses of the kicked beams were dominated by scraping at aperture restrictions,
so there is no straightforward way to determine the contribution of diffusion to those lifetimes.
Nevertheless, the diffusion coefficients and the associated time scales should be useful for 
relative measures of beam growth and particle loss. As an example, it could be useful in IOTA to
quickly distinguish between lattices with different degrees of integrability. 
If echoes can be generated by small amplitude
kicks, then the calculated diffusion coefficients and the time scales would be more representative 
of beam behavior under nominal conditions. Determining the diffusion coefficients may require 
different parameterizations of $D(J)$
at small and large amplitudes, as seen for example in \cite{Chen_92}.

\section{Summary}
In this article, we revisited earlier observations of transverse beam echoes in RHIC
to extract diffusion coefficients from those measurements. 
We considered three models for the action dependence of the diffusion coefficients: 
$D(J)=D_0 + D_1(J/J_0)$, $D(J)=D_0 + D_2(J/J_0)^2$, and 
$D(J)=D_1(J/J_0) + D_2(J/J_0)^2$. 
All three models were found to adequately describe the echo amplitudes measured during
scans of the nonlinear tune shift and the delay between the dipole and quadrupole kicks.
Next, turn by turn data was used to extract both the amplitude and the FWHM of the
pulse width. Here both models with $D_0$ do not describe the data with larger pulse 
widths, so the only model that successfully describes both the amplitude and the FWHM data is the 
$(D_1, D_2)$ model.  We find that $D_1$ is an order of magnitude larger than $D_2$ in
most cases; it increases linearly with the charge while $D_2$ is nearly 
independent
of the charge. Using these charge dependencies, the $(D_1, D_2)$ model also
adequately describes another set of data where the echo amplitudes were measured as
a function of charge. 

These results show that transverse echoes can indeed be used to measure transverse
beam diffusion in existing and future hadron synchrotrons, 
We make some observations on requirements for future measurements. 
The diffusion measurements require good control of several machine and beam parameters
such as the initial dipole kick, the quadrupole kick, machine nonlinearity, tunes and 
beam emittance, to name the most important. Injection oscillations can
strongly influence the echo amplitude and pulse shape, so these need to be controlled
to the extent possible. Alternatively if available, a fast dipole kicker in the ring 
would be preferable to initiate the echo. In such a case, a transverse damper can damp initial oscillations and then be turned off before the dipole kicker is used. While the echo amplitude variation 
with scans of the tune shift and time delay are useful, detailed analysis of the turn by
turn data yields more information. As an example of this, we found that
the FWHM scales quadratically with the charge and therefore is more 
sensitive to intensity changes than the echo amplitude. The proximity of resonances
can also spoil echoes so the tunes and the dipole kick amplitudes need to be chosen
carefully as well. 

\vspace{2em}

\noi {\bf \large Acknowledgments} \newline
Fermilab is operated by Fermi Research Alliance, LLC under U.S. Department of Energy contract No. DE-AC02-07CH11359.
BNL is operated by Brookhaven Science Associates, LLC under U.S. Department of Energy contract No. DE-AC02-98CH10886.

\end{document}